\documentclass[graphics,twocolumn,floatfix,mathbbm,pra,superscriptaddress,aps]{revtex4}
\usepackage{amsmath,amssymb,graphicx,color}
\usepackage{float}
\usepackage{xcolor}
	\newcommand{\Prs}{$\text{Pr}^{3+}\text{:Y}_2\text{SiO}_5$ }
	\newcommand{\Prss}{$\text{Pr}^{3+}\text{:Y}_2\text{SiO}_5$}
	\newcommand{\3}{$^{3+}$}
	
	\newcommand{\mm}[1]{\,\mathrm{#1}}

	\newcommand{\e}{$|e\rangle$}

	\newcommand{\gcc}{$g^{(2)}_{s,i}$\,}
	\newcommand{\gccafc}{$g^{(2)}_{AFC,i}$\,}
	\newcommand{\gc}{$g^{(2)}_{s,i}$}

	\newcommand{\mmu}{\mu\mathrm{m}}

	\definecolor{darkgreen}{RGB}{0,170,50}

\begin{document}

\title{Laser-written integrated platform for quantum storage of heralded single photons}

\author{Alessandro Seri}
\altaffiliation{These authors contributed equally to this paper}
\affiliation{ICFO-Institut de Ciencies Fotoniques, The Barcelona Institute of Technology, 08860 Castelldefels (Barcelona), Spain}

\author{Giacomo Corrielli}
\altaffiliation{These authors contributed equally to this paper}
\affiliation{Istituto di Fotonica e Nanotecnologie - Consiglio Nazionale delle Ricerche and Dipartimento di Fisica - Politecnico di Milano, P.zza Leonardo da Vinci 32, 20133 Milano, Italia}

\author{Dar\'io Lago-Rivera}
\affiliation{ICFO-Institut de Ciencies Fotoniques, The Barcelona Institute of Technology, 08860 Castelldefels (Barcelona), Spain}
\author{Andreas Lenhard}
\affiliation{ICFO-Institut de Ciencies Fotoniques, The Barcelona Institute of Technology, 08860 Castelldefels (Barcelona), Spain}
\author{Hugues de Riedmatten}
\affiliation{ICFO-Institut de Ciencies Fotoniques, The Barcelona Institute of Technology, 08860 Castelldefels (Barcelona), Spain}
\affiliation{ICREA-Instituci\'{o} Catalana de Recerca i Estudis Avan\c cats, 08015 Barcelona, Spain}
\author{Roberto Osellame}
\affiliation{Istituto di Fotonica e Nanotecnologie - Consiglio Nazionale delle Ricerche and Dipartimento di Fisica - Politecnico di Milano, P.zza Leonardo da Vinci 32, 20133 Milano, Italia}
\author{Margherita Mazzera}
\email{margherita.mazzera@icfo.es}
\affiliation{ICFO-Institut de Ciencies Fotoniques, The Barcelona Institute of Technology, 08860 Castelldefels (Barcelona), Spain}

%\date{\today}

\begin{abstract}
Efficient and long-lived interfaces between light and matter are crucial for the development of quantum information technologies. Integrated photonics solutions for quantum storage devices offer improved performances due to light confinement and enable more complex and scalable designs. We demonstrate a novel platform for quantum light storage based on laser written waveguides. The new writing regime adopted allows us to attain waveguides with improved confining capabilities compared to previous demonstrations. We report the first demonstration of single photon storage in laser written waveguides. While we achieve storage efficiencies comparable to those observed in massive samples, the power involved for the memory preparation is strongly reduced, by a factor 100, due to an enhancement of the light-matter interaction of almost one order of magnitude. Moreover, we demonstrate excited state storage times 100 times longer than previous realizations with single photons in integrated quantum memories. Our system promises to effectively fulfill the requirements for efficient and scalable integrated quantum storage devices. 
\end{abstract}

\maketitle

\section{Introduction}
Implementing solid-state quantum storage devices with guided wave optics has several advantages as compactness, scalability, efficiency due to enhanced light-matter interaction, and improved mechanical stability \cite{Orieux2016}. The possibility of connection with other integrated quantum devices, i.e. single photon sources \cite{Vergyris2017}, photonic circuits \cite{pitsios2017photonic}, and detectors \cite{Raffaelli2018}, enables the implementation of complex integrated quantum architectures. In addition, the compatibility of waveguide-based devices with fiber optics opens the way to the interconnection between quantum memories and the current fiber networks, \cite{Bussieres2013} with proven extraordinary telecommunication capabilities. 

Rare earth ion doped crystals, already widely known as long-lived and multiplexed optical memories with quantum storage capabilities \cite{Riedmatten2015}, promise to be excellent systems for the development of on-chip quantum storage devices. The different approaches pursued in this respect include the quantum light storage in Tm$^{3+}$ or Er$^{3+}$ in LiNbO$_3$ waveguides \cite{Saglamyurek2011,Askarani2018} and the storage with optically controlled retrieval of weak coherent states in a nanophotonic crystal cavity in Nd$^{3+}$:YVO$_4$ \cite{Zhong2017}. 
In these remarkable realizations, the ions used do not offer the possibility of storage at the ground state, thus the storage times are inherently limited by the lifetime of the excited state (further shortened in the case of the nanophotonic crystals by the Purcell enhancement in the cavity).
Photon echoes and long spin state lifetimes have been observed in a hybrid integrated system \cite{Marzban2015} composed by a TeO$_2$ slab waveguide deposited onto a \Prs substrate, that potentially enables efficient \cite{Hedges2010} and long lived storage with on-demand read-out \cite{Heinze2013}. This solution, nonetheless, sacrifices the enhancement of the light matter interaction, as the coupling between the guided wave optics, the TeO$_2$  waveguide, and the optically active Pr$^{3+}$ ions happens through evanescent field.

Optical channel waveguides can be also fabricated in the bulk of crystalline substrates by Femtosecond Laser Micromachining (FLM). 
This powerful technique is rapid, cost-effective and features unique three-dimensional fabrication capabilities.
Different writing regimes can be adopted \citep{chen2014optical}. One possibility consists in irradiating the sample at high energy fluence, for producing a pair of damaged material tracks that form the waveguide cladding, and light is guided between them due to a stress-induced positive refractive index change. This kind of structure is called type II waveguide. Another possibility, substantially different from the previous one, consists in writing directly the waveguide core and fabricating a so-called type I waveguide. In this case, a much lower energy fluence is required for a positive refractive index change at the irradiated material volume. Identifying a processing window for operating in the first regime is relatively simple, in fact type II waveguides have been demonstrated in numerous different materials, including rare-earth doped crystals, mainly for integrated laser source applications \cite{tan201070,calmano2011crystalline,grivas2012tunable}. Type II waveguides have also been recently fabricated in \Prs and successfully employed for the coherent storage of classical pulses \cite{corrielli2016integrated}. On the contrary, the fabrication of type I waveguides in crystalline substrates is a very challenging task, since it requires finding a very narrow processing parameters window, if any. So far, this fabrication regime has been demonstrated only in a very limited number of cases \cite{calmano2015crystalline,rodenas2011high,li2017laser}, including lithium niobate\cite{burghoff2007origins,osellame2007femtosecond,thomson2006optical}, potassium dihydrogen phosphate (KDP) \cite{huang2015waveguide} and polycrystalline matrices \cite{rodenas2011direct,macdonald2010ultrafast}.

In the present paper we report on the realization of type I waveguides in a \Prs crystal and we characterize the light guiding properties at 606 nm, where Pr$^{3+}$ features an optical transition of interest for photonic storage purposes. The small dimensions of the waveguide modes guarantee the compatibility with optical fibers and a significant enhancement of the light-matter interaction. Moreover, spectroscopic investigations reveal that the fabrication doesn't affect the optical properties of Pr$^{3+}$ ions in the light guiding region. Finally, to assess the potential of our new platform as quantum memory, we implement quantum storage of heralded single photons. The achieved storage times are 100 times longer than in previous experiments with single photons in waveguides \cite{Saglamyurek2011,Askarani2018}. 

\section{A new waveguide writing regime in \Prs}\label{fes}
\subsection{Fabrication of type I waveguides}
\begin{figure*}[t]
\centering
\includegraphics[width=0.9\textwidth]{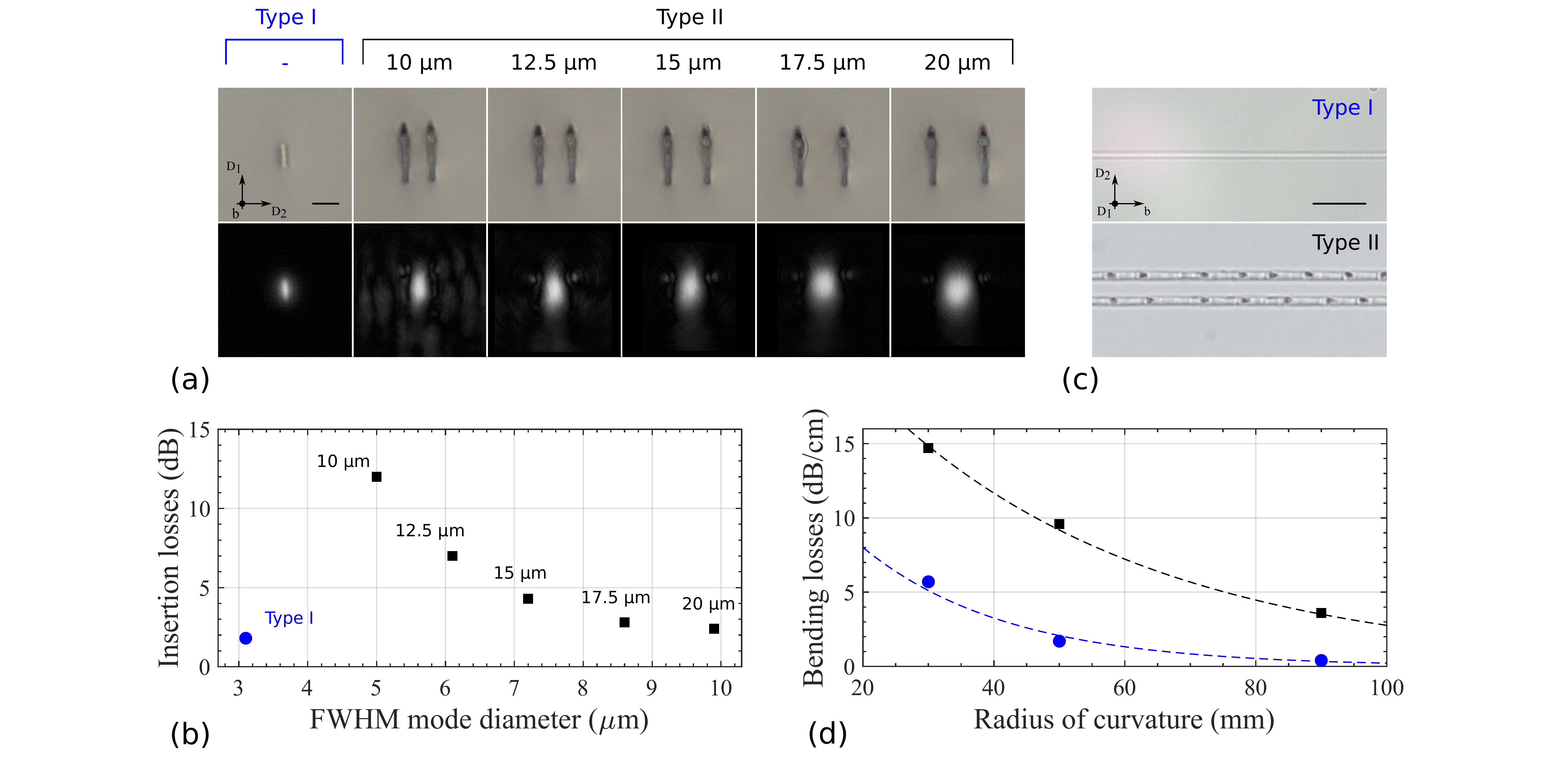}
\caption{(a) Optical microscope image of the waveguide transverse cross section (top row) and guided mode intensity profile (bottom row) of all fabricated waveguides. The numbers above the figures specify the separation between the laser tracks in type II waveguides. Scale-bar is 10 $\mmu$. (b) Insertion losses vs FWHM mode diameter in horizontal direction for type I (blue circle) and type II (black squares) waveguides. (c) Microscope picture of the longitudinal profile of a type I (top) and a type II waveguide with $d$~=~10 $\mmu$ (bottom). Scale-bar is 20 $\mmu$. (d) Bending losses as a function of the radius of curvature for type I (blue circles) and type II (black squares) waveguides. Dashed lines are best exponential fits of experimental data \cite{hunsperger1995integrated}. Error-bars in plots (b) and (d) are smaller than the data markers.}
\label{fig1}
\end{figure*}

Type I waveguides have been directly written by FLM in the volume of a \Prs crystal with a dopants concentration of 0.05\%. We employed the second harmonic at 520 nm of an Yb-based commercial fs-laser source (SPIRIT-1040, Spectra-Physics) which emits a train of ultrashort laser pulses with duration of 350 fs. The waveguides are written in a single transverse scan geometry, with the laser beam propagating along the crystal $\mathrm{D_1}$ axis and the sample translating along the $b$ crystallographic direction (3.7 mm length). The laser pulses are focused 100 $\mmu$ below the sample top surface by means of a 0.6 NA microscope objective. We experimentally found the optimal irradiation conditions for obtaining high quality optical waveguides as 40 nJ pulse energy, 100~$\mmu$/s translation speed, and 20 kHz repetition rate, giving a total fabrication time of $\approx 40$~s for a single waveguide. These waveguides support a single optical mode at the wavelength of 606 nm, polarized along the crystal $\mathrm{D_2}$ axis. An important parameter in the waveguide fabrication is the polarization of the writing laser, which must be linear and perpendicular to the sample translation direction, otherwise no guiding effect is observed at the irradiated lines. The refractive index change of the core with respect to the substrate was estimated numerically from the guided mode profile (according to the method described in ref. \cite{mansour1996improved}) as $\Delta n\approx 1.6\times 10^{-3}$. As type I waveguides in crystals are often thermally unstable \cite{thomson2006optical}, it is worth mentioning that no visible degradation of our type I waveguides in \Prs was observed, after several months of exposure to normal ambient and cryogenic conditions.  

The physical mechanisms that contribute to inducing a type I waveguide in crystals strongly depend on the specific material considered, encompassing the formation of lattice defects and a local change in the material polarizability. Understanding their relative weight is a difficult task. Detailed studies performed on type I waveguides written in LiNbO$_3$ and Nd:YCOB crystals showed that the positive refractive index change in these materials results mainly from a weak lattice distortion and partial ion migration effects taking place at the irradiated area \cite{burghoff2007origins,rodenas2011high}. Such modifications, typically observed in a regime close to the material processing threshold, essentially preserve the bulk properties of the crystal, as demonstrated for type I waveguides in LiNbO$_3$ \cite{osellame2007femtosecond}. Regarding type I waveguides in \Prss, a fundamental explanation of the origin of the positive index change induced by ultrafast laser irradiation is still under investigation.

\subsection{Benchmarking type I vs. type II waveguides}
An experimental characterization was performed for comparing the guiding properties of type I and type II waveguides fabricated in  \Prss. To this purpose, five different type II waveguides were inscribed in the sample employing the same setup described before, but using the laser fundamental harmonic at 1040 nm, a pulse energy of 575 nJ and a scan speed of 60 $\mmu$/s. The five waveguides differ in the separation $d$ between the tracks forming the cladding, ranging from 10 $\mmu$ to 20 $\mmu$. It should be noted that the chosen irradiation parameters provide the best waveguiding properties for all values of $d$. Values of $d \geq 25$ $\mmu$ led to multimode behavior. 

We coupled all waveguides with light at 633 nm from a He-Ne source, polarized along the D$_2$ crystal axis, by means of a plano-convex lens (75 mm focal length, 1 inch diameter). The $1/e^2$ diameter of the laser beam before the lens was 6 mm. We collected the light at the waveguide output by means of a microscope objective (40x, 0.65 numerical aperture). In this way we were able to acquire, with a CCD camera, the normalized mode intensity profile of all waveguides, reported in figure \ref{fig1}(a) together with a microscope picture of their transverse cross sections. In addition, we measured for every waveguide the total insertion losses (IL), defined as the ratio between the light power measured before and after the waveguide. Note that the value of IL includes the waveguide propagation losses due to scattering and absorption, the coupling losses due to mode mismatch at the waveguide input, and the Fresnel losses caused at the crystal/air interface at the waveguide output, where no anti-reflection coating was present. The results of this measurement are shown in fig. \ref{fig1}(b), where we plot the IL values as a function of the horizontal FWHM mode diameter for all waveguides. It is clearly visible that the reduction of $d$ in type II waveguides (black squares) reduces the mode size, but, at the same time, leads to a dramatic increase of IL.  On the other hand, the type I waveguide (blue circle) supports the smallest mode (3.1 $\mmu$~$\times$~5.9~$\mmu$ FWHM diameters) and, simultaneously, exhibits the lowest value of IL, among all waveguides analyzed. This fact is readily explained by looking at the waveguides longitudinal profiles shown in fig. \ref{fig1}(c). The type I waveguide features a smooth and very uniform profile along the propagation direction. The side tracks of type II waveguides, instead, present a rough and less uniform profile, that increases light scattering, especially for small values of $d$. A more detailed analysis of the different contributions to the waveguide IL can be found in the Appendix.

As a further comparison, we measured for the two types of waveguide the additional bending losses (BL), caused by the coupling of light to radiative modes during the propagation in a curved guided path. We thus compared the IL of a straight waveguide with that of a curved one, containing an S-band of known length and fabricated with constant radius of curvature R$_C$. We performed this measurement for values of R$_C$ of 30 mm, 50 mm, and 90 mm, both on the type I and on type II waveguides with $d=20$ $\mmu$, fabricated in a dedicated sample (S-band length = 7 mm, total sample length = 9 mm). The results obtained are shown in fig. \ref{fig1}(d). As expected, the values of BL increase for shorter R$_C$ for both types. However, in type I the measured BL reach particularly low values, and become almost negligible for R$_C >$90 mm. This outcome agrees with the well known fact that BL decrease in more confining waveguides \cite{hunsperger1995integrated}.

This experimental analysis allows us to conclude that type I waveguides in \Prs show better guiding performances than their type II counterparts, both in terms of waveguide losses and in terms of light confinement. Interestingly, the measured values of BL for type I waveguides are compatible with the fabrication of complex integrated devices, i.e. directional couplers or waveguide arrays, as bending radii in the order of tens of mm allow for a flexible engineering of evanescent waveguide coupling, even in samples with a limited length.

\section{Experimental setup}\label{es}
\begin{figure*}[t]
\centerline{\includegraphics[width=1.9\columnwidth]{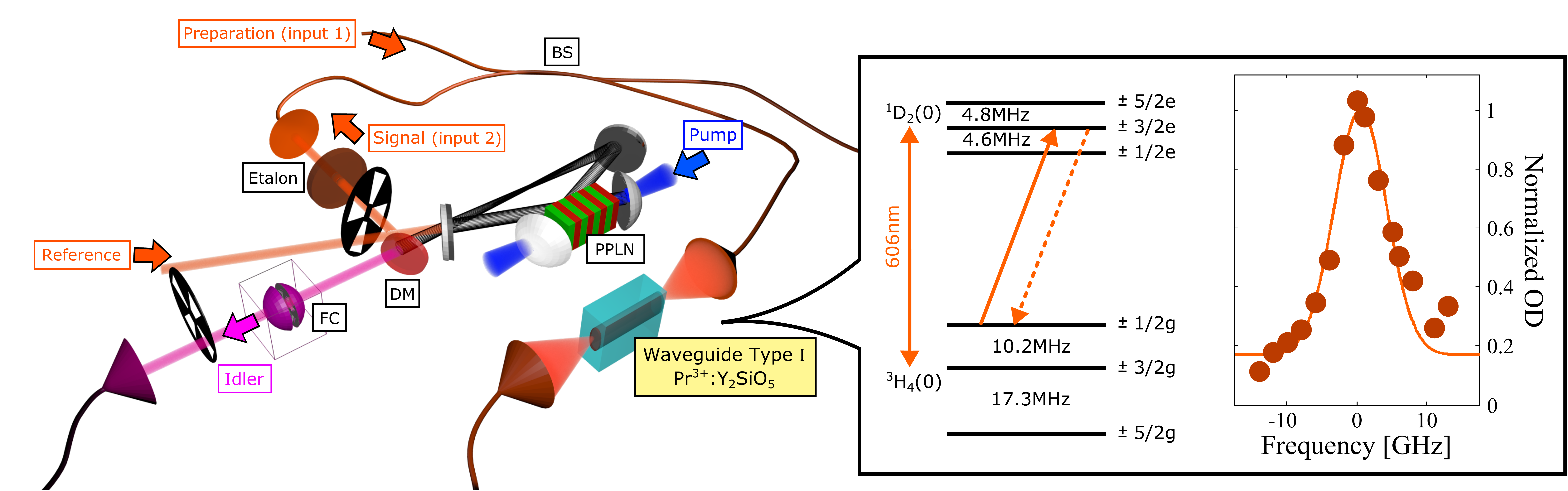}}
\caption{Setup: a pump laser at $426\mm{nm}$ (blue beam) shines a periodically-poled lithium niobate (PPLN) crystal in a bow-tie cavity. Photon pairs (gray beam) are generated by spontaneous parametric down-conversion and divided with a dichroic mirror (DM). The idler photon, at $1436\mm{nm}$ (purple beam), is sent to a filter cavity (FC). The signal photon, at $606\mm{nm}$ (orange beam), passes through an etalon and enters the input 2 of a fiber beam-splitter (BS). Two chopper wheels are used to alternate between locking and measurement period. The signal photon and the preparation light (from input 1 of the BS) emerging from one output of the BS are coupled into the \Prs waveguide. The hyperfine splitting of the first sublevels of the ground $^3H_4$ and the excited $^1D_2$ manifolds of Pr\3 ions in Y$_2$SiO$_5$ is shown in the inset along with the optical absorption spectrum (the data points are experimental values and the solid curve the Gaussian fit). The orange arrows highlight the specific transition chosen for the storage protocol. } 
\label{fig2}
\end{figure*}

Fig. \ref{fig2} represents the experimental setup for the spectroscopy and single photon storage measurements, presented in the next paragraphs, and the hyperfine splitting of the ground and excited crystal field levels of Pr$^{3+}$ involved in the optical transition relevant for the storage (inset). Note that we address only ions in one of the two possible crystallographic sites (site 1).

A CW laser at $606\mm{nm}$ (linewidth 20 kHz) is modulated with acousto-optic modulators (AOMs) in double-pass configuration to produce the preparation pulse sequences. The preparation light is coupled into one input of a fiber beam splitter (BS, input 1 in fig. \ref{fig2}). In input 2 we send the input for the storage, either classical pulses or heralded single photons (see below). One output is sent to an independent optical table where the \Prs crystal is maintained at $3\mm{K}$, while the second output is used as reference.  
The light is coupled into the waveguide using a $75\mm{mm}$ lens, which focuses the beam to a waist  $<10\,\mu$m at the input facet of the crystal. The outcoming light from the waveguide is collected with a $50\mm{mm}$ lens and sent to a detection stage, after a path of about $2\mm{m}$. The detection is implemented with a CCD camera, for imaging and alignment, with a photo-detector, for protocols with classical light, or with a single photon detector (SPD) for experiments with single photons. For autocorrelation measurements a Hanbury Brown-Twiss setup is assembled with fiber beam splitters and additional SPDs. All the experiments are synchronized with the cycle of the cryostat ($1.4\mm{Hz}$). Because of vibrations, the light is efficiently coupled in the waveguide only for less than 300 ms in each cycle. Further details on the setup can be found in the Appendix.

Our heralded single photons are created with a new generation \cite{Maring2018} of the photon pair source described in \cite{Fekete2013,Rielander2016}. It is based on cavity-enhanced type I spontaneous parametric down-conversion (SPDC) in a 2 cm-long periodically-poled lithium niobate (PPLN) crystal (fig. \ref{fig2}). This is placed in a bow-tie cavity (BTC) and pumped with a $426\mm{nm}$ laser. The BTC is in resonance both with the signal photon at $606\mm{nm}$ and its heralding  photon at $1436\mm{nm}$ (idler). The cavity lock exploits the Pound-Drever-Hall technique \cite{Rielander2016} and two mechanical choppers are used to alternate between the locking period and the single-photon measurement. 
The two collinearly-generated photons are separated after the BTC using a dichroic mirror (DM, fig. \ref{fig2}). The signal and idler photons are distributed in 8 spectral modes. The idler photons pass through a home-made Fabry-Perot filter cavity (FC in fig. \ref{fig2}, linewidth $80\mm{MHz}$, FSR=$17\mm{GHz}$), to guarantee single-spectral-mode heralding. They are then coupled into a single mode fiber to an SPD. The $606\mm{nm}$ photons, after passing through an etalon filter, are coupled to a single-mode polarization-maintaining (PM) fiber, and then to the input 2 of the fiber BS. The heralding efficiency of the SPDC source is $\eta_H^{\text{SPDC}} \sim25\%$ after the PM fiber and $\eta_H^{\text{WG}} \sim 7\%$ in front of the waveguide. The loss is due to the BS and to mode-diameter mismatch between the PM fiber and the fiber BS, and could be readily reduced by using a fiber switch.

\section{Spectroscopic and coherence measurements}\label{scm}
The inhomogeneous broadening of the transition of Pr\3 at $606\mm{nm}$ in the waveguide has been estimated from the transmission of an optical pulse through the crystal while tuning the laser frequency. The full-width at half maximum (FWHM) is $\sim 9\mm{GHz}$ in optical depth (OD) (see inset in Fig. \ref{fig2}), in good agreement with bulk samples with similar ion concentrations. Moreover the central frequency did not shift with respect to the bulk. The hyperfine splitting of the levels and the oscillator strength of the transitions, measured following the spectral hole burning experiments described in ref. \cite{Nilsson2004}, are also consistent with those measured in the bulk. The unaltered spectroscopic properties confirm that the fabrication process does not substantially alter the crystalline structure of the modified zone. 

\begin{figure*}[t]
\centerline{\includegraphics[width=2\columnwidth]{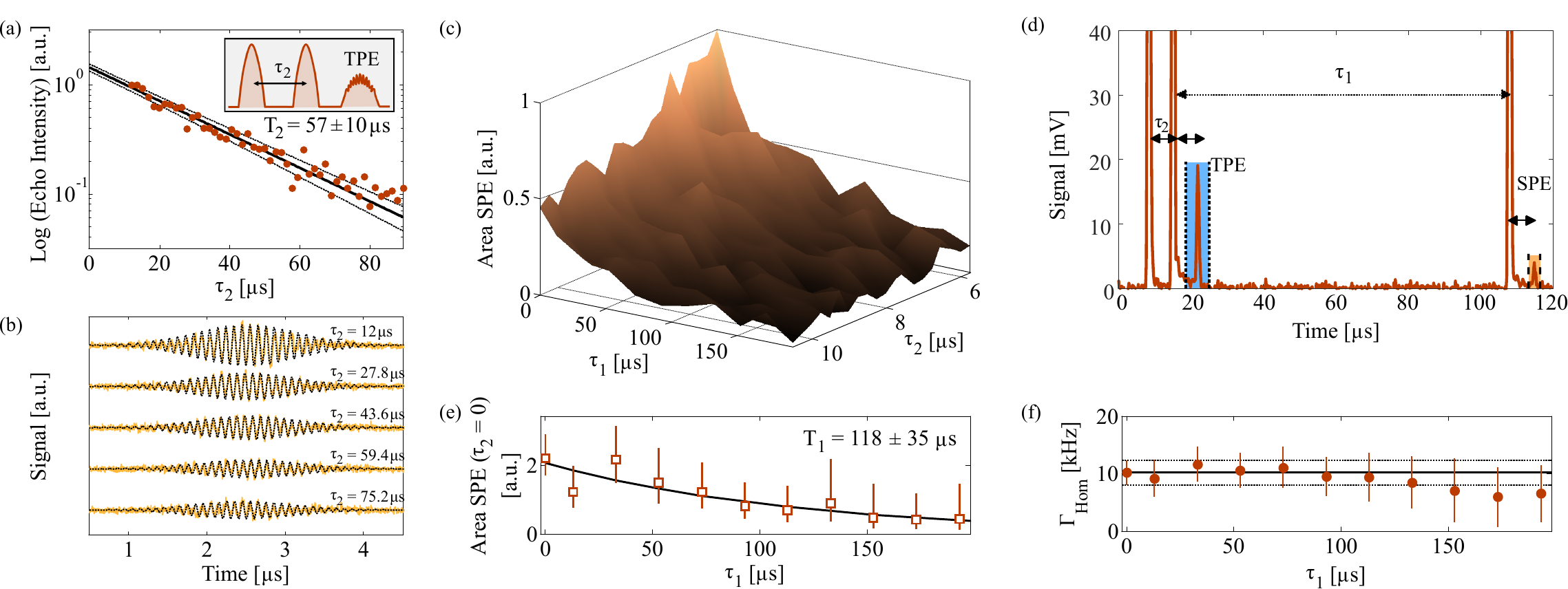}}
\caption{(a) Decay in the echo intensity for a two-photon echo pulse measurement, from which we extract the optical transition coherence time $T_2$ in the type I waveguide. The black solid line is the fit to an exponential decay and the dotted lines indicate the error. The inset shows the temporal sequence used. The time interval between the first and the second pulse is called $\tau_2$ (solid arrow). (b) Examples of heterodyne detected TPE at different times $\tau_2$. The solid orange line is the experimental signal and the dotted black line the sinusoidal fit. The echo intensity is proportional to the square of the oscillation amplitude. (c) Area of the stimulated photon echo (SPE), in arbitrary units, varying the times $\tau_1$ and $\tau_2$; (d) Temporal trace of the SPE process. The SPE signal is highlighted by the orange area. The time interval between the second and third pulse is called $\tau_1$ (dotted arrow); (e) Values of the SPE areas at $\tau_2=0$ (extracted from the exponential decays of the SPE areas over $\tau_2$) plotted as a function of $\tau_1$. From the decay of the SPE vs $\tau_1$ (for $\tau_2=0$) we extract the excited state lifetime of the ions, $T_1$; (f) Homogeneous broadening of the optical transition for increasing $\tau_1$. The values are extracted from the decays of the SPE vs $\tau_2$ for different $\tau_1$ values. The black solid line is the value extracted from the $T_2$ measurement for $\tau_1=0$, the dashed lines indicating the error. } 
\label{fig3}
\end{figure*}

To use the waveguide for quantum memory application the coherence of the ions in the guiding zone should be also maintained. Coherence and lifetime of the optical transition are measured by means of, respectively, two-pulse \cite{Durrant1989} and three-pulse stimulated photon echo \cite{Durrant1991} experiments (TPE and SPE). A typical pulse sequence for the heterodyne detection of the TPE is shown in the inset of Fig. \ref{fig3}a. The echo is detected in form of oscillations on a probe pulse 10 MHz detuned with respect to the excitation and rephasing pulses. From the exponential decay of the echo intensity (proportional to the square of the oscillation amplitude) while increasing the time $\tau_2$ (fig. \ref{fig3}a), we extract the coherence time of the Pr\3 ions in the waveguide. Fig. \ref{fig3}b represents examples of heterodyne detected TPE at different storage times. In our case, probing a single-class absorption feature at the $1/2g-1/2e$ transition, we measure a maximum coherence time $T_2 = 57 \pm 10\, \mu s$. This is only slightly lower than the maximal $T_2$ measured in bulk samples or type II waveguides \cite{corrielli2016integrated}. We note however that the $T_2$ is affected by instantaneous spectral diffusion and that higher values could in principle be achieved by decreasing the average number of atoms excited \cite{corrielli2016integrated}. 

To study the lifetime of the transition and slow dephasing mechanisms, e.g. spectral diffusion, we analyze the SPE measured with direct detection. A typical output of a SPE measurement is shown in fig. \ref{fig3}c while panel d sketches the temporal sequence used. To reduce the errorbar, we perform these measurements with a higher excitation power than that used in the TPE measurement, thus we expect the absolute value of $T_2$ to be lower due to instantaneous spectral diffusion. We measure the decay of the SPE with $\tau_2$ for different values of $\tau_1$. By fitting these decays with exponentials, we extract the expected areas of the $\text{SPE}\, (\tau_1$) for $\tau_2=0$ (fig. \ref{fig3}e). Their decay with $\tau_1$ directly depends on the lifetime of the excited state \e, namely $T_1 = 118 \pm 35\,\mu$s. This value is lower than that obtained from fluorescence measurements (about $160 \,\mu$s) but in good agreement with that measured with SPE in the same bulk sample, $T_1 = 103\pm 11\,\mu$s. In presence of spectral diffusion, the ions could experience frequency shifts induced by flips of the surrounding nuclear spins. This would cause a slow broadening of the ions linewidth, $\Gamma_{hom}$, for increasing $\tau_1$ (\cite{Yano1992}).  As we can clearly see from fig. \ref{fig3}f, $\Gamma_{hom}(\tau_1) = 1/\left[\pi T_2(\tau_1)\right]$ remains consistent over time with the value at $\tau_1 = 0$ (solid line) (about 10 kHz), proving the absence of spectral diffusion in the analyzed timescale and excitation power. The same trend is observed in SPE measurements in the bulk crystal. 

One of the advantages of optical waveguides is the enhanced light-ions interaction due to the strong light confinement. To quantify the strength of this interaction we measure the Rabi frequency $\Omega_R$ of the optical transition by means of optical nutation \cite{Sun2000}. We prepare by optical-pumping a single-class absorption feature on the $1/2g-3/2e$ transition and measure the population inversion time $t_{\pi}$ induced by a long resonant probe pulse. For an optically dense and inhomogeneously broadened ensemble, the Rabi frequency $\Omega_R$ is calculated as $\Omega_R t_\pi = 5.1$ (from eq. 1 in ref. \cite{Sun2000}). We repeat the measurement for several probe powers, $P$. From the linear fit of $\Omega_R$ vs $\sqrt{P}$, we extract $\Omega_R = 2\pi \times 1.75 \mm{MHz}/\sqrt{\text{mW}}$, i.e. an increase of 1.6 with respect to the type II waveguide \cite{corrielli2016integrated} and almost one order of magnitude with respect to the bulk crystal (maintaining the same optics), fully matching with the expected increase due to the stronger light confinement (further details can be found in the Appendix). 

\section{Storage of heralded single photons}\label{shsp}

\subsection{Characterization of the heralded single photons}\label{hsp}
\begin{figure}[t]
\centerline{\includegraphics[width=1\columnwidth]{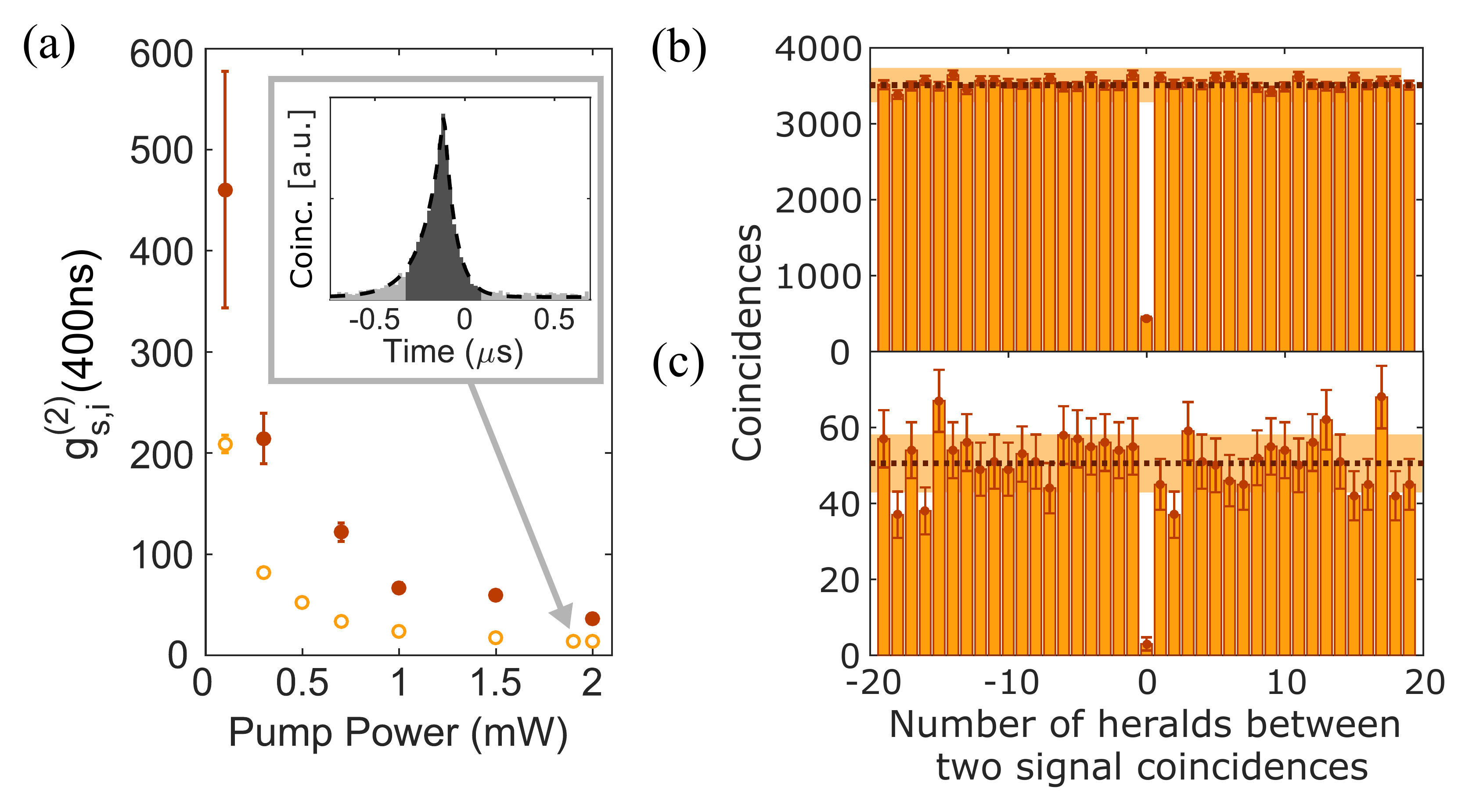}}
\caption{(a) Measured \gcc, for different pump powers, with just the source (empty light dots) and after the {\emph{pit}} (full dark points). The inset represents the time-resolved signal-idler coincidence histogram with just the source: the darker region is the window considered for the calculation of the \gcc ($400\mm{ns}$) and the black dashed line is the temporal fit of the biphoton correlation; (b, c) Coincidences between signal photons splitted with a fiber BS, respectively before and after passing through a \emph{pit} in the waveguide, sorted by the number of heralding photons between pairs of contiguous signal counts.} 
\label{fig4}
\end{figure}

Before storing the signal photons, we measure the properties of the generated heralded single photons (more details can be found in the Appendix).
We build a coincidence histogram using the idler detection as start and the signal photon as stop (inset in fig. \ref{fig4}a). The correlation time of our biphoton is estimated by fitting the histogram \cite{Rielander2016} (black dashed lines) as $121\mm{ns}$, which corresponds to a biphoton linewidth of $\Gamma = 1.8\mm{MHz}$. This is smaller than the hyperfine splitting of the excited state, thus narrow enough to address a single transition of Pr\3 ions. We calculate the normalized second-order cross-correlation function as $g^{(2)}_{s,i}(\Delta t) =p_{s,i}/(p_s\cdot p_i)$, where $p_{s,i}$ is the probability to detect a coincidence in a temporal window $\Delta t$, while $p_s$ ($p_i$) is the probability to detect a signal (idler) count in a temporal window of the same size. The measured \gcc values for a window $\Delta t = 400\mm{ns}$ for different pump powers are plotted in fig. \ref{fig4}a (empty orange circles). The highest value, $g^{(2)}_{s,i}(400\mm{ns}) = 209\pm 9$, is achieved at the lowest measured pump power ($0.1\mm{mW}$), and decreases while increasing the pump power, as expected for a two-mode squeezed state \cite{Sekatski2012}.

We demonstrate the quantum nature of the signal-idler correlations violating the Cauchy-Schwartz (CS) inequality by more than 20 standard deviations (see Appendix for details). The classical bound is set by the parameter $R = {({g^{(2)}_{s,i}) \,^2} \over{g^{(2)}_{s,s}\cdot g^{(2)}_{i,i}}}\le 1$ where $g^{(2)}_{s,s}$ and $g^{(2)}_{i,i}$ are the unconditional auto-correlations of the signal and idler photons, respectively.
%%%%%%%%%%%%% Heralded Autocorrelation 4
To demonstrate the single photon nature of our source, we measure the heralded auto-correlation of the signal photons, $g^{(2)}_{i:s,s}(\Delta t)$. This can be extracted from the histogram of fig. \ref{fig4}b, built like in ref. \cite{Fasel2004, Rielander2016}. We find $g^{(2)}_{i:s,s}(400\mm{ns}) = 0.12\pm0.01$ for a pump power of $1.7\mm{mW}$. This value is considerably lower than the classical bound $g^{(2)}_{i:s,s} \ge 1$ and compatible with the single photon behavior $(g^{(2)}_{i:s,s} \le 0.5)$. See Appendix for more details. 

We then send the signal photons through the waveguide, where we hole-burn a transparency window (\emph{pit}) of $\sim 16\mm{MHz}$ in the inhomogeneously broadened absorption profile of the ions (orange line in the inset of fig. \ref{fig5}a). We measure the \gcc vs pump power sending the signal photons through the \emph{pit} (full brown circles in fig. \ref{fig4}a). The correlations after the \emph{pit} become remarkably higher because the crystal acts as a spectral filter: the \emph{pit} selects a single frequency mode while all the others are absorbed by the Pr$^{3+}$ ions spread over the whole inhomogeneously broadened absorption line \cite{Rielander2014, Zhang2012, Beavan2013}.
We find $ R=524\pm84$ after the \emph{pit} for the highest measured power, violating the CS inequality by more than 6 standard deviations (assuming $g^{(2)}_{s,s}=2$). Furthermore, we measure the $g^{(2)}_{i:s,s}(\Delta t)$ of the signal photons after the \emph{pit}: from the histogram of fig. \ref{fig4}c, we extract $g^{(2)}_{i:s,s}(400\mm{ns}) = 0.06\pm0.04$ (pump power $\sim 1.7\mm{mW}$).

\subsection{Storage of heralded single photons}
The storage protocol that we use is called Atomic Frequency Comb (AFC) \cite{Afzelius2009}: the inhomogeneously broadened absorption of the ions is shaped in a spectral comb with periodicity $\Delta$. When a photon is absorbed by the comb, its state is mapped onto a collective excitation of atoms, described by a collective Dicke state: $\sum_k e^{-i2\pi\delta_kt} c_k|g_1 \cdot\cdot\cdot e_k \cdot\cdot\cdot g_N	\rangle$, where the frequency detuning of the atom $k$ is $\delta_k = m_k\Delta$ (being $m_k$ an integer number). After a first inhomogeneous dephasing, the atoms will collectively rephase at a pre-determined time $\tau = 1/\Delta$, giving rise to a photon re-emission in the forward direction, called AFC echo. 

\begin{figure}[h!]
\centerline{\includegraphics[width=1\columnwidth]{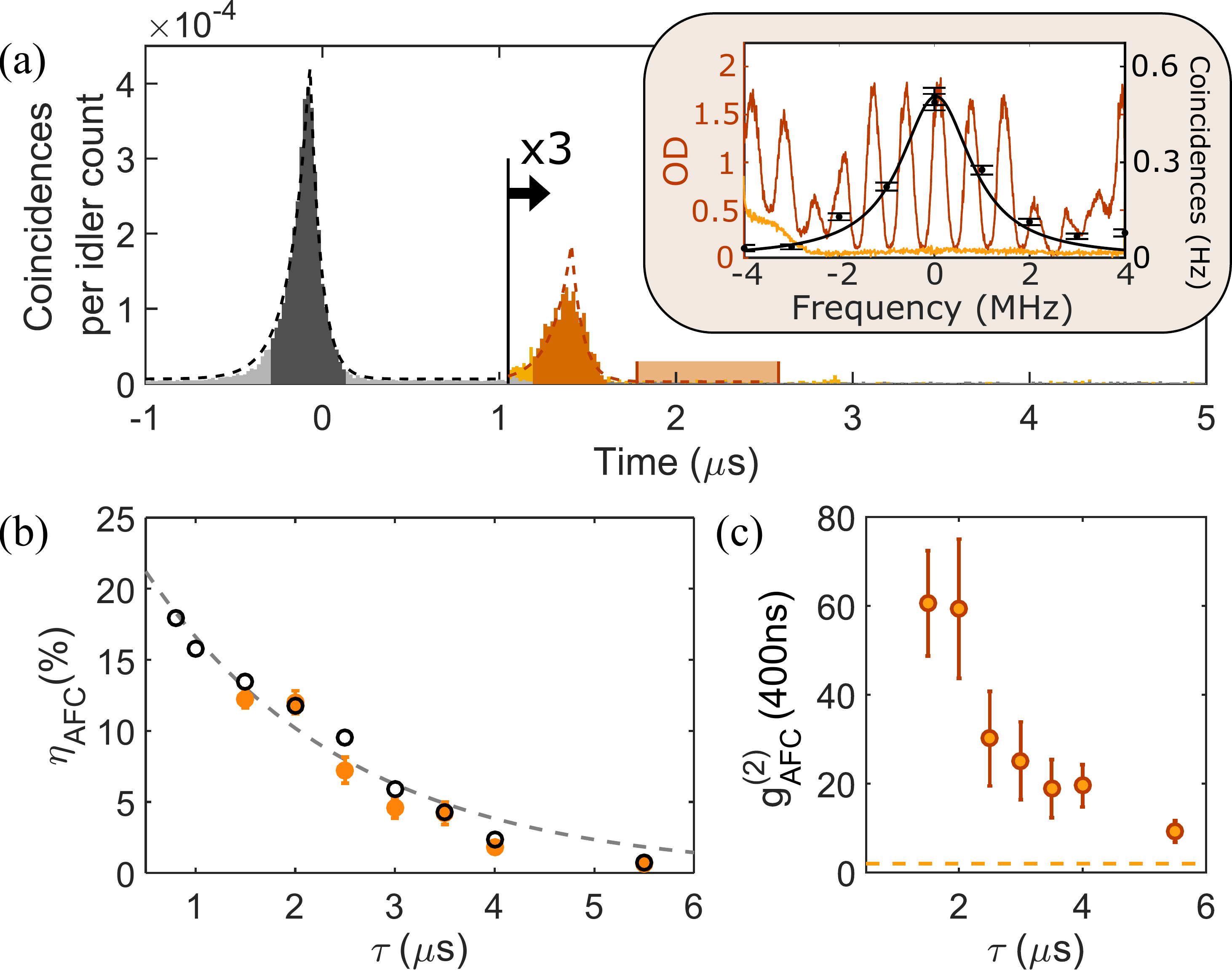}}
\caption{(a) Time-resolved histogram of the idler-signal coincidences for signal photons passing through the \emph{pit} (gray histogram) or the AFC for $\tau=1.5\,\mu$s (orange histogram). The counts in the AFC are multiplied by 3. The darker regions of the peaks show the windows considered for $\eta_{AFC}$ and \gcc calculations. The black and brown dashed lines are the temporal decays of the correlations (inset of fig. \ref{fig4}a), renormalized for losses and efficiencies after the \emph{pit} and the AFC, respectively. The shaded rectangle is the region in which the accidental counts for the AFC echo are measured. The absorption profiles of the \emph{pit} (orange) and the comb (brown) are plotted in the inset, with the spectrum of the single photons (black points and line, see Appendix); (b) Internal storage efficiency $\eta_{AFC}$, at different storage times $\tau$, for single photons (full orange points, error bars account for Poissonian statistics) and classical light (empty black circles); (c) Cross-correlation values between idler photons and stored signal photons, \gccafc, for different $\tau$. The orange dashed line is the classical upper bound, $\sqrt{g^{(2)}_{s,s}\cdot g^{(2)}_{i,i}}=1.58\pm 0.02$ (assuming $g^{(2)}_{s,s}=2$).} 
\label{fig5}
\end{figure}

The sequence of optical pulses used to create the AFC is very similar to the one explained in refs. \cite{Seri2017, Jobez2016}: after creating a $16\mm{MHz}$-wide \emph{pit} to empty the $1/2g$ and $3/2g$ states, we repopulate the $1/2g$ level with a single class of ions (duration $\sim80\mm{ms}$). Then we send a series of pulses, resonant with the $1/2g-3/2e$ transition, whose Fourier transform is the AFC that we want to create (duration $\sim35\mm{ms}$). 
The generated feature, for a storage time of $1.5\,\mu$s, is plotted as a brown line in the inset of fig. \ref{fig5}a. Note that the spectrum of the input photons (black line in the inset of fig. \ref{fig5}a) matches perfectly with the AFC (see Appendix for details on the bi-photon spectrum). Thanks to the enhanced light-ions interaction, the maximum power that we inject in the waveguide during the AFC creation is $\sim 100\,\mu$W, i.e. two orders of magnitude lower than what is usually necessary in bulk, in agreement with the $\Omega_R$ increase. The measurement is performed in the remaining time ($\sim150\mm{ms}$), resulting in a duty cycle for the AFC storage of $\sim 21\%$ (accounting for the duty cycle due to the cryostat vibrations). 
When a herald is detected, an AOM in front of the pump laser is closed to reduce the noise coming from the source during the AFC echo retrieval (see Appendix). The minimum response time of this gate is $t_\mathrm{P}^\mathrm{off} = 1.2\,\mu$s, which limits our minimum storage time to $\tau = 1.5\,\mu$s. For longer storage times, $t_\mathrm{P}^\mathrm{off}$ is delayed in order to maintain $\tau - t_\mathrm{P}^\mathrm{off} = 0.3\,\mu$s. The AFC echo for a storage time $\tau=1.5\,\mu$s is shown in fig. \ref{fig5}a (orange trace). The exponential fit of the biphoton temporal decay, measured with only the source (black dashed line of the inset in fig. \ref{fig4}a), is plotted on top of the input (black dashed line) and the AFC echo (brown dashed line), renormalized for the different count-rates. 
Note that the linewidth of the photons, both after the \emph{pit} and emitted by the comb, remains unchanged, confirming that there is no mismatch between the width of the comb and the spectral profile of the photons. Before and after each storage experiment, we measure the coincidences between the idler and the signal photons after the \emph{pit}. To account for fluctuations in power, we consider the average of the two as our reference input (gray peak in fig. \ref{fig5}a). We evaluate the storage efficiency by integrating the counts of the histogram in a $400\mm{ns}$ window centered at the AFC echo (dark orange region at about $1.5\,\mu$s, fig. \ref{fig5}a) divided by the counts inside a $400\mm{ns}$ window centered at the input (dark gray region about $0\,\mu$s, fig. \ref{fig5}a), the latter normalized by the transmission through the \emph{pit} ($85\%$, see inset in fig. \ref{fig5}a). The resulting efficiency is the internal efficiency of the process, $\eta_{AFC}$. The total efficiency of our device, i.e. the ratio between the output signal and the input signal before entering the waveguide, is calculated multiplying $\eta_{AFC}$ by the coupling into the waveguide ($\sim 40\%$). We perform storage experiments with an average pump power of $1.7\mm{mW}$. The internal efficiencies for different storage times are shown in fig. \ref{fig5}b (full orange points). For comparison we measure the internal efficiency of our memory with classical pulses (empty black circles in fig. \ref{fig5}b), using the same comb preparation sequences, showing a good overlap between the quantum and the classical regimes. The efficiency decrease for increasing $\tau$ is fitted with an exponential decay, $e^{-4/(T_2^*\Delta)}$ \cite{Jobez2016}, from which we extract the effective coherence time of our storage protocol, $T_2^* = 8\,\mu$s, much smaller than the $T_2$ (see section \ref{scm}). This suggests that our storage time is at the moment limited by technical issues and not yet by the coherence time $T_2$ of the Pr\3 in the waveguide. We estimate that a factor 2 could be given by instantaneous spectral diffusion because, due to the time limit imposed by the cryostat cycle, we implement the optical pumping for the AFC preparation with relatively high power. The remaining mismatch between $T_2^*$ and $T_2$ is likely due to the finite laser linewidth (see section \ref{es}) and power broadening.

The \gccafc of the AFC echo is measured similarly to the one of the input: we find $p_{s,i}$ integrating the counts in a window of $400\mm{ns}$ centered at the AFC echo (the same region considered for $\eta_{AFC}$); $p_s\cdot p_i$ is measured integrating the accidentals coincidences after the AFC echo up to the last stored noise count, $t_\mathrm{P}^\mathrm{off} + \tau$ (light orange rectangle in fig. \ref{fig5}a), renormalized to a $400\mm{ns}$ window. Fig. \ref{fig5}c shows the \gccafc values for AFC echoes measured at different $\tau$. The cross-correlation increases after the storage up to $61\pm12$ for a storage time of $1.5\mu$s, with respect to $36\pm3$ after the \emph{pit}. This could be explained by the presence of broadband noise from the SPDC source, which is not in resonance with the inhomogeneous absorption line of the Pr\3 ions. Such noise would not be present in the temporal mode of the AFC echo, where the pump is gated off \cite{Rielander2014}. The value of the \gccafc should remain constant for different $\tau$, as the storage efficiency is the same for the AFC echo and for the stored noise. But for longer storage times, as $\eta_{AFC}$ decreases, our signal-to-noise ratio is limited by the background noise. Nevertheless, the \gccafc remains higher than the classical bound (orange dashed line) for all the measured storage times up to $\tau=5.5\,\mu$s, for which we violate the CS inequality with $R=34 \pm 18$ (above the classical bound by almost 2 standard deviations), effectively demonstrating the longest quantum storage in an integrated solid-state optical memory (100 times longer than the previous demonstrations of single photon storage in waveguides \cite{Saglamyurek2011, Askarani2018}). Moreover, thanks to the convenient energy level scheme, our system enables the full spin-wave AFC storage, thus giving access to both longer storage times and on-demand read-out \cite{Gundogan2015,Jobez2015,Seri2017}.

\section{Conclusion and Outlook}

In the present work we propose a new platform for the implementation of integrated quantum storage devices. We demonstrated the generation of type I waveguides in a \Prs crystal using fs-laser micromachining in a new writing regime. We showed that the fabrication of type I waveguides preserves the measured spectroscopic properties of Pr$^{3+}$. We implemented a quantum storage protocol for heralded single photons, observing high non-classical correlations for storage times 100 times longer than in previous waveguide demonstrations. 
The use of type I waveguides in \Prs gives several advantages with respect to type II ones. In type I waveguides, the guided mode is in general sensibly smaller than what obtainable in a type II waveguide with comparable losses. This fact yields an enhancement in the interaction of the guided light with the rare earth dopants. Moreover, the very good mode matching between type I waveguides and standard single mode optical fibers would allow us to glue directly the \Prs samples to fiber patch cords with low coupling losses, simplifying sensibly the procedure of light coupling inside the cryostat and avoiding the temporal constraints on the photon storage given by the cryostat vibrations. High-quality type I waveguides will also allow to produce linear cavities with high quality factors by directly writing Bragg gratings superimposed to the waveguide \cite{marshall2008directly}. In addition, type I waveguides in \Prs could also be easily interfaced with laser written optical circuits in glass, potentially opening the way to the realization of integrated hybrid glass/crystal platforms \cite{atzeni2018integrated} embedding quantum memories. Remarkably, taking advantage of the intrinsic three-dimensional capabilities of FLM, one can envision high spatial multiplexing by an efficient exploitation of the substrate volume, with matrices of quantum memories interconnected to linear fiber arrays by glass circuits. Finally, for this kind of waveguides, the absence of lateral damage tracks (present in the type II counterpart) enables greater freedom in engineering the evanescent coupling of light between different waveguides. This, together with the tighter bend radii achievable in type I waveguides, permits to easily inscribe optical circuits in \Prs crystals embedding directional couplers and other integrated optics devices, for performing complex tasks besides quantum light storage, fully on chip.

\section{Acknowledgment}
We acknowledge financial support from the European Research Council (ERC) through the Advanced Grant project CAPABLE (grant agreement n. 742745). ICFO further acknowledges financial support by the Spanish Ministry of Economy and Competitiveness (MINECO) and Fondo Europeo de Desarrollo Regional (FEDER) (FIS2015-69535-R), by MINECO Severo Ochoa through grant SEV-2015-0522 and the PhD fellowship program (for A.S.), by Fundaci\'o Cellex, and by CERCA Programme/Generalitat de Catalunya.

\section{Appendix}\label{SuppWGI}

In this Appendix we report the details about the characterization of type I and type II waveguides, the experimental setup and the measurements performed to characterize the light matter interaction and the heralded single photon source used in the main paper.

\subsection{Waveguides characterization}\label{waveguides}
\begin{table*}[]
\centering
\begin{tabular}{|c|c|c|c|c|c|c|c|}
\hline
 \textbf{WG Type} & \textbf{d ($\boldsymbol{\mu}$m)} & \textbf{$\Delta\mathbf{_H}$ ($\boldsymbol{\mu}$m)} & \textbf{$\Delta\mathbf{_V}$ ($\boldsymbol{\mu}$m)} & \textbf{IL (dB)} & \textbf{CL (dB)} & \textbf{FL (dB)} & \textbf{PL (dB/cm)} \\ \hline
I & - & 3.1 & 5.9 & 1.8 & 0.84 & 0.37 & 1.6 \\ \hline
II & 10 & 5.0 & 12.1 & 12.0 & 1.09 & 0.37 & 28.5 \\ \hline
II & 12.5 & 6.1 & 12.5 & 7.0 & 1.12 & 0.37 & 14.9 \\ \hline
II & 15 & 7.2 & 12.5 & 4.3 & 1.21 & 0.37 & 7.4 \\ \hline
II & 17.5 & 8.6 & 12.0 & 2.8 & 1.31 & 0.37 & 3.0 \\ \hline
II & 20 & 9.9 & 12.5 & 2.4 & 1.68 & 0.37 & 0.9 \\ \hline
\end{tabular}
\caption{Summary of the results obtained from the characterization of the type I and II laser written waveguides presented in the main text. $d$ is the distance between the side tracks of type II waveguides, $\Delta_{H/V}$ are the FWHM horizontal/vertical dimensions of the guided mode, IL stands for Insertion Losses, CL stands for Coupling Losses, FL stands for Fresnel Losses and PL stands for Propagation Losses.}\label{Tab1}
\end{table*}
In table \ref{Tab1} we list the values of the FWHM mode dimensions in the horizontal ($\Delta_H$) and vertical ($\Delta_V$) directions and the different contributions to losses for the type I and II waveguides presented in the main text.

From the values of $\Delta_{H,V}$ it is possible to appreciate that the type I waveguide supports a sensibly smaller mode than the type II ones. In addition, it is possible to see that $\Delta_H$ in type II waveguides strongly depends upon the distance $d$ between the tracks, while $\Delta_V$ is almost unaffected by this parameter. This is consistent with the fact that the horizontal confinement in type II waveguides is provided by the high refractive index change of the core with respect to the lateral tracks, where the material becomes amorphous. The vertical confinement, instead, is given by the weak refractive index increase that occurs in the region between the tracks due to stress effects \cite{Gorelik2003}, and it is only barely influenced by $d$. In fact, the value of $\Delta_V$ is mainly determined by the vertical height of the laser-induced tracks, which measures approximately 20 $\mmu$. It is worth highlighting that the vertical extension of the tracks is a hardly controllable parameter during the fabrication of type II waveguides. It is, in fact, much longer than the confocal parameter of the focused laser beam inside the crystal, (around 4 $\mmu$ in our case) and it is mainly determined by non-linear propagation effects (e.g. self focusing) caused by the very high optical power involved. A slight elongation in the vertical direction is visible also in type I waveguides, but in this case, as the writing power is sensibly smaller, it can be attributed mainly to the spherical aberrations of the focusing objective.

We performed the Insertion Losses (IL) measurement by coupling He-Ne light at $\lambda$=633~nm into the waveguides using a plano-convex lens with focal length $f$=75 mm, and collecting the light at the waveguide output with a microscope objective (40x, 0.65 NA). The laser beam impinging the lens was expanded to a $1/e^2$ diameter D$_0$=6~mm (circular cross section). This produces a FWHM focal spot size at the waveguide input facet equal to $\Delta_0=5.9$ $\mmu$. This value can be calculated by using the formula $\Delta_0\approx 2.36\cdot \lambda f/(\pi D_0)$, obtained from the theory of Gaussian beam focussing \cite{Svelto1998}. The value of IL in dB is calculated by the formula $\mathrm{IL=-10\log_{10}(P_{out}/P_{in})}$, where $\mathrm{P_{in}}$ and $\mathrm{P_{out}}$ are the light powers measured before and after the waveguides, respectively. This value comprises three distinct contributions: the coupling losses (CL) that arise from the mismatch between the focal spot and the waveguide modes, the Fresnel losses (FL) that are caused by reflection at the air/crystal interfaces, and the propagation losses (PL) caused by scattering of light from waveguide imperfections and by absorption of \Prs at 633 nm along propagation. We estimated the values of CL by calculating numerically the superposition integral $\eta$ between the field distribution at the focal spot $\mathrm{E_{IN}}=\exp[-1.39(x^2+y^2)/\Delta_0^2]$ and that of the waveguides mode $\mathrm{E_{WG}}=\exp[-1.39(x^2/\Delta_H^2+y^2/\Delta_V^2)]$, as defined in \cite{Osellame2004}. The value reported in table \ref{Tab1} is then calculated as $\mathrm{CL}=-10\log_{10}(\eta)$. The single-interface FL are calculated as $\mathrm{FL}=-10\log_{10}(1-r)$, where $r$ is the Fresnel reflection coefficient obtained considering $n_{\mathrm{c}}=1.8$ for the \Prs refractive index. Finally, PL are calculated as $\mathrm{PL=(IL-CL-FL)/L_c}$, where $\mathrm{L_c}$=3.7 mm is the crystal length. In this formula FL are counted only once, because an anti-reflection coating is present on the crystal input facet. It is worth noting that, despite the significant difference in mode dimensions between the waveguides, the variation in CL is rather limited when compared to the variation observed for IL. As a consequence, the IL trend presented in fig. 1(b) of the main text can be almost entirely ascribed to the different values of PL. In the case of type type II waveguides PL are highly affected by the scattering due to the roughness of the side tracks, and this effect becomes more and more relevant for decreasing values of $d$. The quasi-periodic damages that affect type II waveguides (visible as black spots in fig. 1(c) of the main text) along the writing direction can be explained by thermal accumulation processes that cause quasi-periodic micro-explosions in the material. They are particularly relevant for type II waveguides where the pulse energy is much higher, while they are not present in type I waveguides. A similar phenomenon has been already observed and explained for FLM in fused silica \cite{Richter2013}.

\subsection{Experimental setup}\label{exp}

The experimental setup for the spectroscopic characterization of the waveguides and single photon storage measurements is sketched in fig. 2 of the main text. We hereby provide a more detailed description. 
Our laser source at $606\mm{nm}$ is a Toptica DL SHG pro, frequency stabilized using the Pound-Drever-Hall technique to a home-made Fabry-Perot cavity placed in a vacuum chamber. We estimate the final laser linewidth to be about 20 kHz. The CW laser light is modulated in amplitude and frequency with acousto-optic modulators (AOMs) in double-pass configuration, driven by an arbitrary waveform generator (Signadyne). The preparation light is coupled into one input of a fiber beam splitter (BS, input 1 in fig. 2 in the main text). In input 2 we send the input for the storage (either classical pulses or heralded single photons, see below). One output is sent to an independent optical table where the \Prs crystal (Scientific Materials) with laser written waveguides is maintained at $3\mm{K}$ (in a He closed cycle cryocooler, Oxford Instrument), while the second output is used as reference.  
The light is coupled into the waveguide using a $75\mm{mm}$ lens, which focuses the beam to a waist  $<10\,\mu$m at the input facet of the crystal. The out coming light from the waveguide is collected with a $50\mm{mm}$ lens and sent to a detection stage, after a path of about $2\mm{m}$. It is worth noting that both lenses are placed outside of the cryostat chamber, thus making the alignment of the whole system rather challenging. The detection is implemented with a CCD camera, for imaging and alignment, with a photo-detector, for protocols with classical light, or with a single photon detector (SPD) for experiments with single photons. 

All the experiments are synchronized with the cycle of the cryostat ($1.4\mm{Hz}$) to reduce the effect of the mechanical vibrations. Two mechanical shutters are installed in the setup during the single photon measurements: one, in front of the SPD, remains closed during the preparation period; the second, in anti-phase with the first one, is installed before the input 1 of the fiber BS, and remains closed during the single photon measurement period, blocking the leakage from the preparation AOM.
The SPD is a Laser Components detector with 50\% detection efficiency and $10\mm{Hz}$ dark-count rate. For unconditional and heralded auto-correlation measurements the waveguide output is split with a fiber BS and an Excelitas SPD with 50\% detection efficiency and $50\mm{Hz}$ dark-count rate is connected to the second output.

Our heralded single photons are generated with a photon pair source (fig. 2 of main text) inspired by the source described in \cite{Rielander2016}. It is based on cavity-enhanced spontaneous parametric down-conversion process (SPDC) in a 2 cm-long type I periodically-poled lithium niobate (PPLN). The non-linear crystal is pumped with a $426\mm{nm}$ laser (Toptica TA SHG) to produce signal photons at $606\mm{nm}$ and idler photons at $1436\mm{nm}$. The crystal is placed inside a bow-tie cavity (BTC) with a free spectral range (FSR) of $261 \mm{MHz}$ to enhance the generation of the photon pairs at the frequencies of the cavity modes. The BTC is maintained resonant to both the signal and its heralding idler photon at $1436\mm{nm}$ with a double lock system. First of all, the cavity length is locked using the Pound-Drever-Hall technique to a reference beam at $606\mm{nm}$ derived from the main memory preparation laser \cite{Rielander2016}. Then, a classical beam at $1436\mm{nm}$, generated as frequency difference (DFG) of the pump and the reference beam at $606\mm{nm}$, is used to ensure the maximum transmission of the idler photons through the BTC. The lock of the DFG signal is operated by acting on the $426\mm{nm}$ pump frequency. Two mechanical choppers are used to alternate between the locking period and the single-photon measurement. The BTC, besides enhancing the non-linear process at the two resonant wavelengths, is meant to generate ultra narrow-band photons \cite{Rielander2016}. 
A consequence of the double lock is the clustering effect given by the different FSR at the signal and idler frequencies \cite{Pomarico2012}. This results in a spectral output of the SPDC source consisting of a main cluster with only $8$ effective spectral modes separated by $261 \mm{MHz}$. 
The photons of the pair are generated collinearly and separated after the BTC using a dichroic mirror (DM, fig. 2 in the main text). The idler photon passes through a home-made Fabry-Perot filter cavity (FC in fig. 2 of the main text, linewidth $80\mm{MHz}$, FSR=$17\mm{GHz}$), to guarantee a single-spectral-mode heralding. It is then coupled into a single mode fiber to an SPD (ID230, IDQuantique), with $10\%$ efficiency and 10 Hz of dark-count rate. For the auto-correlation measurement of the idler photons, we use a second SPD (ID220, IDQuantique), with $10\%$ efficiency and 400 Hz of dark-count rate. The $606\mm{nm}$ photons then pass through an etalon (linewidth $4.25\mm{GHz}$, FSR=$100\mm{GHz}$) that suppresses the side clusters. Finally they are coupled in a single-mode polarization-maintaining (PM) fiber and then to the input 2 of the fiber BS. The heralding efficiency of the SPDC source is $\eta_H^{\text{SPDC}} \sim25\%$ after the PM fiber and $\eta_H^{\text{WG}} \sim 7\%$ in front of the waveguide.

\subsection{Light-matter interaction}\label{wr}
\begin{figure}[h]
\centerline{\includegraphics[width=1\columnwidth]{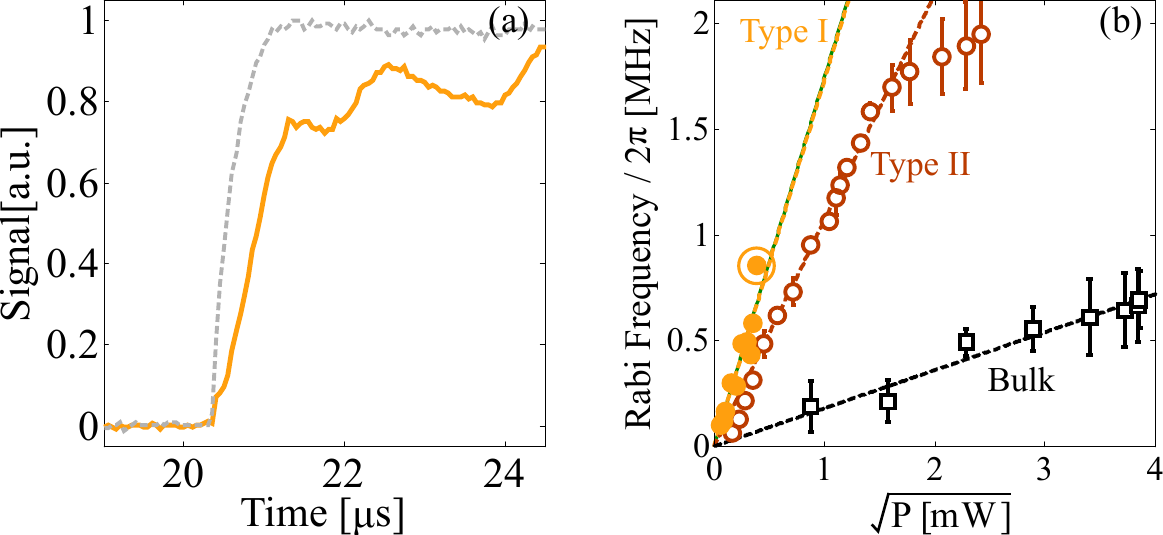}}
\caption{(a) The solid orange trace is the measured intensity of a long light pulse, $P = 0.51\mathrm{mW}$, transmitted by a single-class absorption feature on the $1/2g-3/2e$ transition. The gray dotted line is the reference pulse before entering the crystal. (b) Rabi frequency as a function of the pulse power as calculated from optical nutation measurements performed on the waveguide type I (filled orange circles), compared with the $\Omega_R$ measured in a longer bulk sample (empty black squares) containing also a  type II waveguide (empty brown circles) \cite{corrielli2016integrated}. The dashed lines are the linear fits of the experimental data. The green solid line is the expected behavior for the type I waveguide from the slope of the type II fit, scaled according to the different diameters. The circled data point in panel (b) refers to the pulse reported in panel (a). }
\label{rabi}
\end{figure}

The interaction between light and matter is highly enhanced in the waveguide, thanks to the high confinement of the light  through the whole length of the crystal. We quantify the strength of this interaction measuring the Rabi frequency $\Omega_R$, by means of optical nutation \cite{Sun2000}: we prepare by optical-pumping a single-class absorption feature on the $1/2g-3/2e$ transition and measure the population inversion time $t_{\pi}$ induced by a long resonant probe pulse (grey dashed trace in fig. \ref{rabi}(a)). For a Gaussian beam in an optically dense and inhomogeneously broadened ensemble, the Rabi frequency $\Omega_R$ is defined as $\Omega_R t_\pi = 5.1$ \cite{Sun2000}. 

We repeat the measurement for several probe powers, $P$ (orange filled points in  fig. \ref{rabi}(b)). We show in  fig. \ref{rabi}(b), for comparison, the measured $\Omega_R$ for different $P$ in a waveguide type II and in bulk (respectively brown empty circles and black empty squares), from our previous work in a longer sample \cite{corrielli2016integrated}, being the dotted lines fits of the data. From the linear fit of $\Omega_R$ vs $\sqrt{P}$ for the type I (orange dotted line), we extract $\Omega_R = 2\pi \times 1.75 \mm{MHz}/\sqrt{\text{mW}}$. This value agrees quite well with that calculated from the dipole moment of the investigated transition ($1.45\times 10^{-32}$ Cm \cite{Nilsson2005}), i.e. $2\pi \times 1.6 \mm{MHz}/\sqrt{\text{mW}}$. For this calculation we consider a Gaussian mode with the average FWHM diameter measured at $606\mm{nm}$ with the same setup used for the optical nutation measurements. In this condition, the measured diameters (4.5 $\mmu$ and 7.6 $\mmu$ in the horizontal and vertical directions, respectively) are slightly different than those quoted in section \ref{waveguides}, as the two setups differ in many aspects, e.g. the light wavelength, the objective and CCD camera. The Rabi frequency measured for the type I waveguides features an increase of 1.6 with respect to the previously measured type II waveguide and almost one order of magnitude with respect to the bulk crystal \cite{corrielli2016integrated}. This result fully matches with the expected increase due to the stronger light confinement, green solid line below the fit, which has been calculated from the fit of the type II (brown dotted line), renormalized for the different mode diameter in the waveguide type I.

\subsection{Characterization of the heralded single photons}

%%%%%%%% Linewidth 4
Before sending the signal photons through the fiber BS and to the memory setup, we directly connect the output of the source to the SPD, to measure the properties of the generated heralded single photons. We record the detection times of both signal and idler photons with a fast time-stamping electronics (Signadyne). 
%% figure 4
\begin{figure}[h]
\centerline{\includegraphics[width=1\columnwidth]{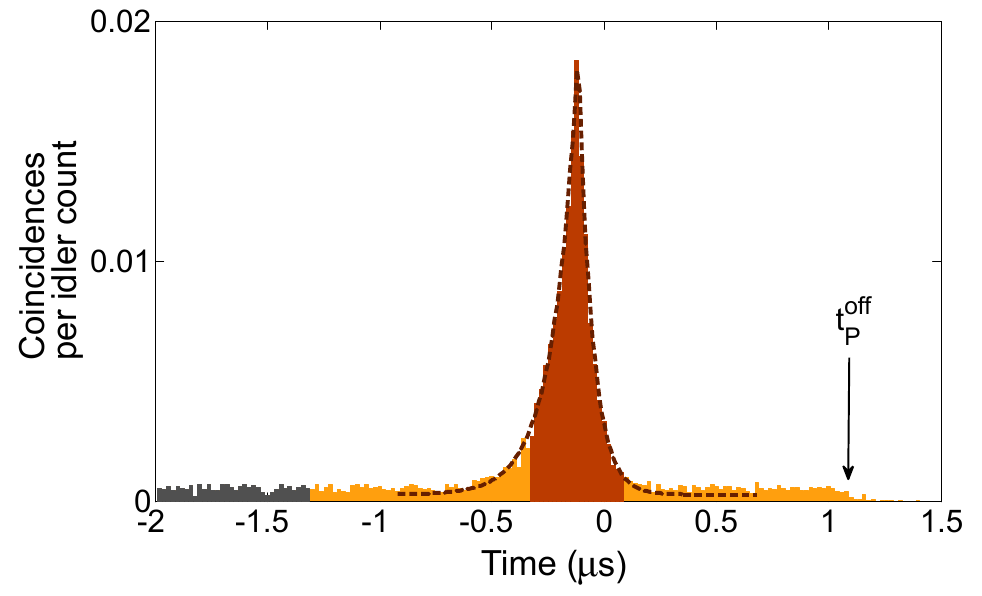}}
\caption{Time-resolved coincidence histogram between the two photons, the darker orange region being the signal-of-interest and the gray region on the left being the accidental coincidences considered for the calculation of the \gcc ($400\mm{ns}$ window). The black dashed line is the temporal fit of the biphoton correlation. The arrow points at the gating off of the pump laser to the photon pair source. } 
\label{sfig1}
\end{figure}

Then we build a time-resolved coincidence histogram, using the idler detections as start and the signal photons as stop (fig. \ref{sfig1}). The correlation time of our biphoton can be estimated by fitting the histogram with two exponential decays \cite{Rielander2016} (black dotted lines on top of the histogram). The two decay times are different due to the different losses experienced in the cavity by the signal and the idler photons, generated at widely different frequencies. From the right (left) decay we can extract the linewidth of the signal (idler) photons to be $\Gamma_s = 2.5\mm{MHz}$ ($\Gamma_i = 1.4\mm{MHz}$). The resulting biphoton linewidth is $\Gamma = 1.8\mm{MHz}$ in FWHM (equivalent to a coherence time of $121\mm{ns}$), narrow enough to address a single transition of Pr\3 ions.

\begin{figure}[h!]
\centerline{\includegraphics[width=1\columnwidth]{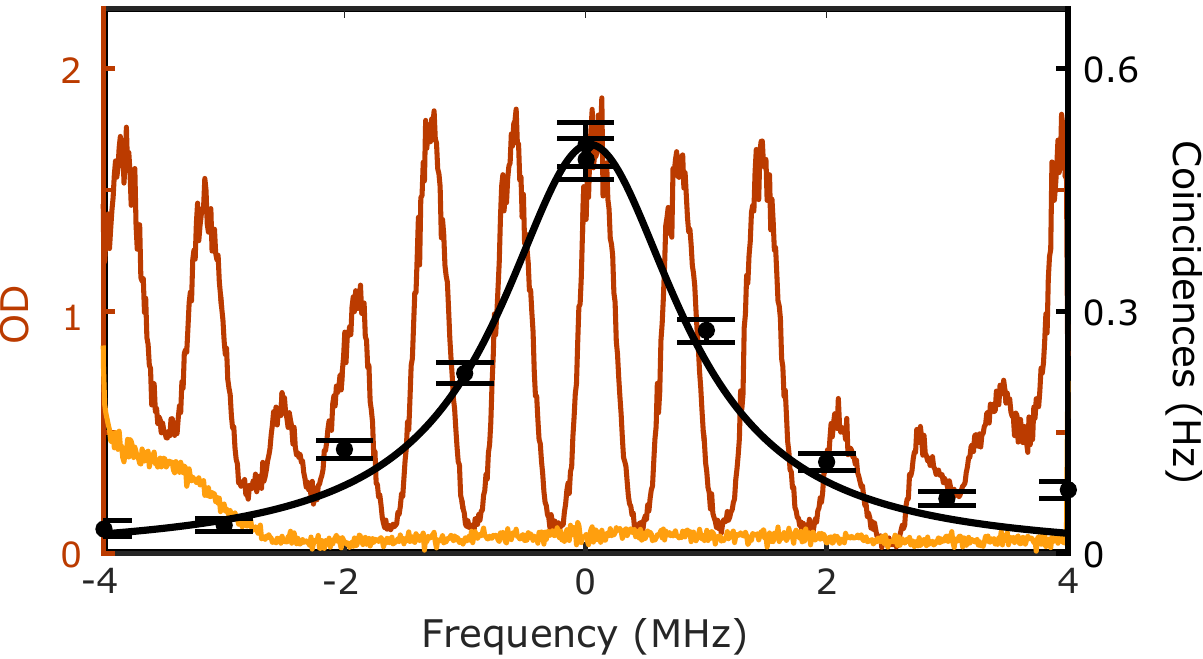}}
\caption{Absorption profiles of the \emph{pit} (orange) and the comb (brown).  The black line represents the spectral shape of the heralded single photons, a Lorentzian function of $1.8\mm{MHz}$ bandwidth, convoluted with the trace of the \emph{pit} used as tunable filter. The black dots are the coincidence rate of the heralded signal photons passing through the waveguide used as a tunable filter centered at different frequencies.} 
\label{sfig2}
\end{figure}

%%%%%%%% Linewidth 5
To confirm the linewidth of the correlations, we use the crystal as a tunable ultranarrow spectral filter, following a measurement described in \cite{Seri2017}: we hole-burn a transparency window (\emph{pit}), centered each time at a different frequency around the signal photons, and we analyze the change in the coincidence rate after it. To prepare a \emph{pit} we shine the crystal with strong pulses at the desired frequency and with a frequency chirp dependent on the width of the \emph{pit} that we want to create, in this case $\sim 1.6\mm{MHz}$. 
The coupling optics is free-space and outside the cryostat (see sect. \ref{exp}). Inside the cryostat is only the waveguide which moves periodically with injection of compressed liquid He in the cooling device (cooling period $707\mm{ms}$), thus the light is effectively coupled in the waveguide for a time $<300\mm{ms}$. Having fiber coupled samples would relax the time limitation and enable better optical pumping.
We \emph{reset} the absorption spectrum and prepare the \emph{pit} at the beginning of each cycle, limiting our measurement duty cycle to $\sim 30\%$. Having fiber-coupled devices would allow us to increase significantly the duty cycle of the experiment. The coincidence rates for \emph{pit}s prepared at different frequencies are plotted in the inset of figure 5 of the main paper (black dots), which is reported here (fig. \ref{sfig2}). The black line, a Lorentzian function with a linewidth of $\Gamma = 1.8\mm{MHz}$ in FWHM, convoluted with the spectral trace of the \emph{pit}, matches with the measured data. This confirms the linewidth estimated from the coincidence histogram (fig. \ref{sfig1}).

%%%%%%%%% Second order cross-correlation 4
We quantify the non-classicality of the photon-pair correlations measuring the normalized second-order cross-correlation function: $g^{(2)}_{s,i}(\Delta t) =p_{s,i}/(p_s\cdot p_i)$, where $p_{s,i}$ is the probability to detect a coincidence in a temporal window $\Delta t$, while $p_s$ ($p_i$) is the probability to detect a signal (idler) count in a temporal window of the same size. The \gcc is extracted from the coincidence histogram of figure \ref{sfig1}, by integrating the coincidence counts in a window $\Delta t$ around the coincidence peak (our signal of interest $p_{s,i}$, dark orange region of the histogram) and dividing them by the coincidences outside of it (accidental coincidences between uncorrelated photons or detector dark counts, $p_s\cdot p_i$, gray region of the histogram) \cite{Rielander2016}, renormalized to a window $\Delta t$. We integrate the counts over a window $\Delta t = 400\mm{ns}$. The measured \gcc values for different pump powers are plotted in figure 4a of the main paper (empty orange circles).

This measurement alone does not demonstrate the quantum nature of our correlations, which can be shown by violating the Cauchy-Schwartz (CS) inequality. The classical bound is given by the parameter $R = {({g^{(2)}_{s,i}) \,^2} \over{g^{(2)}_{s,s}\cdot g^{(2)}_{i,i}}}\le 1$, where $g^{(2)}_{s,s}$ ($g^{(2)}_{i,i}$) is the auto-correlation of the signal (idler) photons. We measure the unconditional auto-correlation of the signal (idler) photon by splitting its path with a fiber BS and detecting the two outputs with two different SPDs. From the resulting coincidence histogram we extract the auto-correlation value (similarly to the cross-correlation measurement). Using an integration window $\Delta t = 400\mm{ns}$ for comparison with the \gc, we find $g^{(2)}_{s,s}(400\mm{ns}) = 1.051 \pm 0.002$ and $g^{(2)}_{i,i}(400\mm{ns}) = 1.25\pm 0.03$. The expected auto-correlation for an ideal state generated by a photon source with thermal statistics is $g^{(2)\,th}_{x,x}(0) = 2$ \cite{Tapster1998}. As the signal photons, measured without any filter cavity, are multi-mode and knowing that we have a number of frequency modes N = 8, we expect for the signal photons $g^{(2)\,th}_{s,s} (0)= 1+1/N = 1.12$ \cite{McNeilGardiner1983}. Anyway the auto-correlation values of both idler and signal photons are lower than expected because we measured them in a $400\mm{ns}$ window (instead of extracting their values in the 0-point) and because the noise generated by the source reduces the auto-correlation value (a detailed discussion can be found in \cite{Rielander2016}). 

Using these values, we find for the lowest pump power $R  = (3.3\pm0.3)\times 10^4$, surpassing the classical bound by more than 10 standard deviations. Even for the highest pump power ($2\mm{mW}$), where the measured cross-correlation is $g^{(2)}_{s,i}(400\mm{ns}) = 13.8\pm 0.3$, we find $R = 145\pm 6$, which violates the CS inequality by more than 20 standard deviations, thanks to the better statistics.

%%%%%%%%%%%%% Heralded Autocorrelation 4
To demonstrate the single photon nature of our source, we measure the heralded auto-correlation $g^{(2)}_{i:s,s}(\Delta t)$: the auto-correlation of the signal is measured conditioned on the detection of an heralding in the same integration window $\Delta t$. The histogram of figure 4b of the main paper, inspired from \cite{Fasel2004}, is built as follows: for each count in the idler detector we look for detection events in the two signal detectors, happening in the same temporal window ($\Delta t =400\mm{ns}$ around the heralding). If there is a count in one of the two signal lists, we look for the closest event in the other one. The triple-coincidences are then sorted depending on the number of heralding events between each two contiguous signal detections. The events in which a coincidence between the two signal detectors is heralded by the same idler count are plotted in the bin 0. The ratio between the counts in bin 0 and the average of the other bins (black dotted line in figure 4b of the main paper) gives the heralded auto-correlation, which we measure to be $g^{(2)}_{i:s,s}(400\mm{ns}) = 0.12\pm0.01$ (pump power $\sim 1.7\mm{mW}$). This value is considerably lower than the classical bound $g^{(2)}_{i:s,s} \ge 1$ and compatible with the single photon behavior $(g^{(2)}_{i:s,s} \le 0.5)$.

\subsection{Characterization of the heralded single photons after the \emph{pit}}\label{shsp}
We connect the signal photons fiber output to the input 2 of the fiber BS and we couple them into the waveguide, as sketched in the setup (figure 2 of the main paper).

%%%%%%%%% Second order cross-correlation 5
We hole-burn a \emph{pit} of $\sim 16\mm{MHz}$ in the inhomogeneously broadened absorption profile of the ions (the spectral trace of the \emph{pit} in OD is plotted as an orange line in figure \ref{sfig2}). We measure the \gcc between the heralding photons and the signal photons passing through the \emph{pit} similarly to the previous section. The results, measured for different powers in the same range of the measurement with just the source, are plotted in figure 4a of the main paper (full brown circles). The non-classicality of the correlations after the \emph{pit} is remarkably higher than the one of the source alone. For example, the \gcc at the highest pump power ($2\mm{mW}$) is found to be $36\pm3$, being almost three times higher than the \gcc of the source alone. 
This can be explained by a spectral filtering effect of the \emph{pit}. The spectral modes in the signal arm that do not have an heralding photon and are not filtered by the etalon (see section \ref{exp}) are in fact absorbed by the atoms outside of the \emph{pit}. 
Due to the low count-rate, caused by transmission losses as well as short measurement duty-cycle, we do not have enough statistics in the auto-correlation measurement of the signal after the \emph{pit} (an exploratory trace more than $20\mm{h}$ long does not show any bunching). We consider the conservative value of 2 for an ideal two-mode squeezed state for the signal \cite{Tapster1998} and the measured value only for the idler photons,  leading to a classical bound of $\sqrt{g^{(2)}_{s,s}\cdot g^{(2)}_{i,i}}=1.58\pm 0.02$. Even with this assumptions we find $ R=524\pm84$ after the \emph{pit} for the highest measured power, violating the CS inequality by more than 6 standard deviations. 

%%%%%%%%%%%%% Heralded Autocorrelation 5
Finally, we measure the heralded auto-correlation of the signal photons after the \emph{pit}: the signals are sent through the crystal, then split with a fiber BS and detected with two different SPDs. From the post-processed histogram (fig. 4c of the main paper) we extract $g^{(2)}_{i:s,s}(400\mm{ns}) = 0.06\pm0.04$ (pump power $\sim 1.7\mm{mW}$).
We know that the $g^{(2)}_{i:s,s}$ is inversely proportional to \gcc for two-mode squeezed states and low pump powers \cite{Chou2004}. This is verified in our measurement, in fact the heralded auto-correlation is lower after the \emph{pit}, where the cross-correlation is higher. For a pump power of $1.7\mm{mW}$, using $g^{(2)}_{s,s}=2$ and the measured value for $g^{(2)}_{i,i}$, we find $g^{(2)\,\text{th}}_{i:s,s} = g^{(2)}_{s,s} \cdot g^{(2)}_{i,i} /g^{(2)}_{s,i} \sim 0.06$ which matches with the measured heralded auto-correlation.

\addcontentsline{toc}{chapter}{Bibliography}
{\bibliographystyle{osajnl}}

\end{document}